\documentclass[twocolumn,showpacs,preprintnumbers,amsmath,amssymb,prl]{revtex4}
\usepackage{graphicx}% Include figure files
\usepackage{dcolumn}% Align table columns on decimal point
\usepackage{bm}% bold math

\begin{document}

%\newcommand{\half}{\mbox{$\frac{1}{2}$}}
%\newcommand{\dyad}[1]{\mbox{\boldmath $#1$}}
%\newcommand{\ddt}{\mbox{$\frac{d}{dt}$}}

% \preprint{}

\title{Bell's Theorem: A New Derivation That Preserves Heisenberg and Locality  }

\author{Michael Clover}
\email{michael.r.clover@saic.com}
\affiliation{%
Science Applications International Corporation\\
San Diego, CA \\
}

%\date{24 May 2004, last revised }
\date{\today}
\begin{abstract}
{By implicitly assuming that all measurements occur simultaneously, Bell's Theorem only applied to local theories that violated Heisenberg's Uncertainty Principle.  By explicitly  introducing time into our derivation of Bell's theorem, an extra term related to the time-ordering of actual measurements is found to augment ({\it i.e.} weaken) the upper bound of the inequality.   Since the same locality assumptions hold for this rederivation as for the original,  we conclude that only {\em classical} measurement-order independent local hidden variable theories are constrained by Bell's inequality;  time dependent, non-classical local theories  ({\it i.e.} theories respecting Heisenberg's Uncertainty Principle) can  satisfy this new bound while exceeding Bell's limit.  Unconditional nonlocality is only expected to occur with Bell parameters between $2\sqrt{2}$ and $4$. This weakening of Bell's inequality is seen for the quantum Bell operator (squared) as an extra term involving the commutators of {\em local} measurement operators.  We note that a factorizable second-quantized wavefunction can reproduce experimental measurements; because such wavefunctions allow local de Broglie-Bohm hidden variable modelling, we have another indication that violation of {\em Bell's} inequality does not require an acceptance of non-locality. }
\end{abstract}

% PACS, the Physics and Astronomy  Classification Scheme.
\maketitle

\noindent

\section{Introduction}

John Bell, in his book, ``Speakable and Unspeakable in Quantum Mechanics''~\cite{B71}  showed that all prior attempts to ``prove'' the completeness of QM made invalid assumptions.  He also showed that the  interpretation offered by de Broglie  and later expanded upon by Bohm~\cite{B52} evaded all those ``proofs'', but when applied to two particles involved a troubling feature:  nonlocality or ``action at a distance.'' 
With the realization that the de Broglie-Bohm (dBB) interpretation resulted in nonlocal forces between particles, Bell developed a mathematical relationship that is taken  to show that  nonlocality is an essential aspect of reality, or at least an essential part of hidden variable theories.    In the last 35-40 years, experiments ({\it e.g.}~\cite{OU88, ADR82, WZ98}) have universally discovered that his bound is violated.

We show that Bell's analysis implicitly assumed a compatibility or simultaneity of measurements in his analysis.  This goes beyond assuming the simultaneous existence of hidden variables to assuming their simultaneous measureability, and this means his derivation denies  Heisenberg's Uncertainty Principle (HUP).  Our derivation removes that assumption by explicitly accounting for the fact that each measurement has to occur at a separate time and ends with a weaker inequality that is not violated by Quantum Mechanics, experiment, nor local non-classical hidden variable theories.

\subsection{Bell's Derivation of his Theorem}

Bell's derivation  of the Bell/CHSH inequality~\cite{B71} starts by calculating the difference of two (theoretical) averages or correlation coefficients,
\begin{eqnarray}
\langle AB\rangle - \langle AB'\rangle &\equiv & \int d\lambda \rho(\lambda)A(a,\lambda)B(b,\lambda)  \nonumber \\
 & &- \int d\lambda \rho(\lambda)A(a,\lambda)B(b',\lambda) \ ,\label{eq:identity} 
\end{eqnarray}
where  $A, B, B' = \pm1$ are the {\em results} of measurements and depend on the orientation of various filters ($a,b$ or $b'$) and may also depend  on hidden variables, $\lambda$, which are assumed to have some distribution, $\rho(\lambda)$. Experimental averages have a similar form, for example, $\langle AB\rangle \equiv N^{-1}\sum_i A_i(a)B_i(b)$, and provide a better notation if we wish to apply this theorem to, say, the Copenhagen Interpretation of quantum mechanics, where $\rho(\lambda) = \delta(\lambda-\Psi)$, {\it i.e.} where the measurements only depend on the wavefunction.  

  The  derivation begins by pulling the integral out of the difference,
  \begin{eqnarray}
  \langle AB\rangle - \langle AB'\rangle & = \langle AB -  AB'\rangle \ , \label{eq:counterfactual}
  \end{eqnarray}
 followed by further mathematical manipulations, finally leading to
 \begin{eqnarray}
|\langle AB\rangle - \langle AB'\rangle| + |\langle A'B'\rangle + \langle A'B\rangle|& \le & 2 \ , \label{eq:bellimit}
\end{eqnarray}
the famous Bell inequality. 

 The act of undistributing the integral  at equation~\ref{eq:counterfactual}  requires us to assume that $B(b,\lambda)$ measurement result is known at the same (timeless) moment that $B(b',\lambda)$ measurement result is known,  the theoretical equivalent of assuming that Heisenberg's Uncertainty Principle doesn't apply to these hidden variables or wavefunctions ({\it i.e.} that $\sigma_z$ and $\sigma_x$ can both be {\em known} simultaneously).  This undistribution has also been termed ``counterfactual'' because experimentally, Bob cannot make both measurements at the same time  -- simultaneous magnetic fields at zero and ninety degrees compose a single field at 45 degrees, {\it etc}.  

%Moreover, since each correlation function can assume values between $-1$ and $+1$, we would like to understand why experiments to date, while measuring Bell parameters (LHS of equation~\ref{eq:bellimit}) in excess of $2$,  are bounded by $2\sqrt{2}$, instead of exhausting the causality limit of $4$.

\section{Bell's Theorem forced to  correspond with Reality\label{sec:belldata}}

We  re-derive Bell's theorem consistent with the Uncertainty Principle   by introducing  a measure of  time into the derivation,  imagining that the four experiments ($\langle AB\rangle , \langle AB'\rangle, \langle A'B'\rangle , \langle A'B\rangle$) are measured sequentially  at times $t_1,\  t_2,\  t_3$ and $ t_4$, respectively\footnote{Modern experiments make individual measurements in a random order, but the essential point is that they are not at the {\em identical} time; there is no loss of generality in the temporal binning used here.}. %end footnote
We will make the same assumption of locality as Bell, that Alice's result, $A$, is only a function of her  setting, $a$, and is independent of Bob's setting, $b$ ({\it i.e.} $A = A(a,\lambda) \ne A(a,b,\lambda)$) and {\it vice versa}.  We differ from  Bell in that we write $A(a(t), \lambda(t))$ in general.  The detector settings, $a(t),\  b(t)$ will each  be constant for the periods of time corresponding to the different correlation measurements, so that  we will write $a'_2$ during the second time interval, and $b_3$ during the third time interval, {\it etc}. 

For typographical convenience, we will assume that any theoretical averages that would have been written as $ \int d\lambda \rho(\lambda)A(a,\lambda)B(b,\lambda)  $ can be converted into a normalized sum over (an arbitrarily  large number of) events, $ N^{-1}\sum_{i=1}^{N} A_i(a, \lambda_i)B_i(b,\lambda_i) $, by ensuring that $\lambda_i$ occurs with a frequency proportional to $\rho(\lambda)$.
 
Beginning with this kind of data model, we can write,
\begin{eqnarray}
\langle AB\rangle_1 - \langle AB'\rangle_2 
&=& N_1^{-1}\sum_{i_1=1}^{N_1} A(a_1,\lambda_{i_1})B(b_1,\lambda_{i_1}) \nonumber \\
 & & -  N_2^{-1}\sum_{i_2=1}^{N_2} A(a_2,\lambda_{i_2}) B(b'_2,\lambda_{i_2})  \ , \nonumber \\
 &\equiv & N_1^{-1}\sum_{i_1=1}^{N_1} A_1B_1  -  N_2^{-1}\sum_{i_2=1}^{N_2} A_2 B'_2  \ , \nonumber 
\end{eqnarray}
where  we have introduced the further abbreviation $A(a'_n, \lambda_{i_n}) \equiv A'_n$, {\it etc}.  We now  assume  a  reordering of  the  elements within each ensemble so that the hidden variables {\em exactly} correspond to each other (or at least arbitrarily closely), allowing us to associate a particular element from one ensemble with a particular  element from another  \footnote{Instead of regarding this as two perfectly matched (``cloned'') ensembles, one could instead imagine this as a single ensemble with the same particle undergoing two consecutive measurements, which has it's own experimental implications.}. %end footnote
Completing Bell's first step (absorbing factors of $N^{-1}$ into the summation sign),
 \begin{eqnarray}
\langle AB\rangle_1 - \langle AB'\rangle_2 
&=&  \sum_{i=1}^{N} A_1B_1   - A_2 B'_2 
  =  \langle AB -  AB'\rangle  \ ,  \nonumber 
\end{eqnarray}
%\newpage
\noindent
allows us to add and subtract terms.   Thus,
\begin{eqnarray}
 \langle AB -  AB'\rangle 
 &=&  \sum [A_1B_1 
           \pm A_1B_1A'_3B'_3  \mp A_1B_1A'_3B'_3\nonumber \\ 
   & &   \ \  \ \    - A_2B'_2 
           \mp A_2B'_2A'_4B_4 \pm A_2B'_2A'_4 B_4] \ , \nonumber
 \end{eqnarray}
factoring terms, 
\begin{eqnarray}
 \langle AB -  AB'\rangle &=&  \sum A_1B_1 [1 \pm A'_3B'_3] %\nonumber \\
          - A_2 B'_2 [1 \pm A'_4B_4] \  \nonumber \\
       & &     \mp \sum A_1B_1A'_3B'_3   
            \pm \sum A_2 B'_2A'_4B_4 \ ,\nonumber \\
&=&  \mbox{Bell's terms (but with subscripts)}  \nonumber \\
        & & \     \mp \sum [A_1 A'_3 B_1 B'_3- A_2 A'_4 B'_2 B_4 ] \ . \nonumber 
\end{eqnarray}
Adding and subtracting another term and refactoring,
\begin{eqnarray}
\langle AB\rangle_1 - \langle AB'\rangle_2 
   &=&  \mbox{Bell's terms}  \nonumber \\     
            & & \     \mp \sum A_1A'_3[B_1B'_3-B'_2B_4] \nonumber \\
             & &       \mp \sum[A_1A'_3 -A_2A'_4]B'_2B_4 \ , \nonumber  
\end{eqnarray}
so that by  taking the absolute value of the left and right hand sides ($|A|\le 1, |B|\le 1$), we have
\begin{eqnarray}
\left| \langle AB\rangle_1 - \langle AB'\rangle_2 \right |
   &\le &  1 \pm \langle A'B\rangle_4  +1 \pm \langle A'B'\rangle_3   \nonumber \\
            & & \ \     +  \sum \left| [B_1B'_3-B'_2B_4] \right|  \nonumber \\
            & &  \  \   +  \sum \left| [A_1A'_3 -A_2A'_4] \right|  \ . \nonumber  
\end{eqnarray}
If individual terms in either of the last two sums are negative, we can replace any $|x-y|$ with $(x-y) + 2(y-x)$, collecting the  latter (a minority of terms) into $|...|$.  In that case, our inequality becomes
\begin{eqnarray}
\left| \langle AB\rangle_1 - \langle AB'\rangle_2 \right |
   &\le &  1 \pm \langle A'B\rangle_4  +1 \pm \langle A'B'\rangle_3   \nonumber \\
            & & \ \    + \left| \sum  [B_1B'_3-B'_2B_4] \right|  + | ... |    \nonumber \\
            & & \ \    +  \left| \sum  [A_1A'_3 -A_2A'_4] \right|  + | ... | \ , \nonumber  
\end{eqnarray}
where the  last term is identically  zero since it is the difference of a correlation function, $\langle A(a,t)A(a',t+\Delta t)\rangle$, with itself.  Skipping some algebra that Bell skipped, and introducing a factor, $f\ge 1$, to account for the $|...|$ terms, we get:
 \begin{eqnarray}
|\langle AB\rangle_1 - \langle AB'\rangle_2| &+&  |\langle A'B'\rangle_3 + \langle A'B\rangle_4|  \nonumber \\
   & \le &  2    + f  \left| \sum [B_1B'_3-B'_2B_4] \right|  \ , \label{eq:gallupslimit}  
\end{eqnarray} 
where the  last term  is  related to whether  $b$ is measured earlier or later than  $b'$ in the constructed correlation coefficients.  {\em If} the order of measurements is irrelevant, we recover Bell's limit of $2$.

   The notion of ``earlier'' and  ``later'' would not make much sense if it only referred to elements from two independent ensembles -- Bertie$_3$'s answer is presumably independent of Bernie$_1$'s answer, and neither has anything to do with Barney$_4$'s or Bob$_2$'s.  However, because we reordered the elements of the ensembles, it is the case that Bernie$_1$ is an {\em identical} clone with Bernie$_2$,  Bernie$_3$ and Bernie$_4$ --  the  same hidden variables, the same names -- the same everything except wall-clock time and detector setting.  The new term in equation~\ref{eq:gallupslimit} can  therefore be interpreted as cumulating the difference between making the $b'$ measurement before and after the $b$ measurement on the {\em same} element of an ensemble at {\em different} times\footnote{Different temporal assignments would allow us to derive an alternative equation with, for example,  $\left| \sum [A_1A'_3-A'_2A_4] \right|$ on the right hand side of equation~\ref{eq:gallupslimit}.}.%end footnote
   
 Our earlier reordering of elements in the ensembles required us to assume that $\lambda_1$ ``matched'' $\lambda_2$; our derivation now presents an alternative.  If the hidden variables are time independent, then $\lambda_3$ will also be numerically identical to  $\lambda_1$ (and $\lambda_4 = \lambda_2$) and it would seem reasonable that both correlation functions would be numerically identical and that Bell's inequality would be satisfied.  If  the hidden variables are dynamic, and if we take seriously the clone-equivalence of Bernie$_3$ and Bernie$_1$\footnote{Everything in this derivation is also consistent with a single particle encountering a follow-on detector in a tandem experimental configuration.}, then the numerical value of  $\lambda_3$ before the $b'$ measurement is made should  be whatever value $\lambda_1$ had after the $b$ measurement was made (and similarly for $\lambda_4$'s evolution from $\lambda_2$).
 
It is not enough that the hidden variables be time dependent;  their dynamical behavior must be such that measuring the $b$-$b'$ correlation in opposite order gives different correlation coefficients, and we shall term those dynamic hidden variable theories that are sensitive to such conditions, {\em non-classical}   (or quantum) HVT's.

Experimentally, if Bob, for example, had two extra polarizing beam splitters (PBS), both oriented at $b'$, one viewing the output of the original PBS's $b$ face, the other viewing the output of the $b_\perp$ face, one could imagine cumulating the $\langle B_1 B'_3\rangle$ term in equation~\ref{eq:gallupslimit}.  When the original PBS is oriented to $b'$, the two tandem PBS's would have to be reset to $b$, and then $\langle B_2' B_4\rangle$ could be measured.  In this manner, one might be able to measure all the terms of equation~\ref{eq:gallupslimit} in two experiments instead of four.

 A  violation of this new  inequality (which has {\em not} been seen)  might be evidence for nonlocality.

\section{Bell's Theorem for Operators}

 The violations of Bell's inequality has led some  authors ({\it e.g.}~\cite{L87,R03, dBMR99}) to construct a quantum mechanical Bell ``operator'',
and to show it to satisfy the identity,
 \begin{eqnarray}
\hat{S}_{Bell}^2 & = &  \left( \hat{A}\hat{B}+\hat{A'}\hat{B} + \hat{A}\hat{B'} - \hat{A'}\hat{B'}\right)^2 \nonumber \\
 & \equiv & 4\hat{I}  - [\hat{A},\hat{A'}] [\hat{B},\hat{B'}] \ , \label{eq:bellsquare}
\end{eqnarray}
on the assumption that the operators are normalized ($\hat{A}^2 = \hat{B}^2 = \hat{I}$) and local ($[\hat{A},\hat{B}] =0$).  If $A$ and $A'$ are interchanged, this expression compares with the LHS of equation~\ref{eq:gallupslimit}, ignoring absolute values, and if $[\hat{A},\hat{A'}]=0$ or $ [\hat{B},\hat{B'}]=0$, we recover the quantum  version of Bell's inequality for Heisenberg violating operators. It has also been shown~\cite{mrc2} that  for EPR-Bell experiments, the operators that measure the projection of states along $(a, a')$, $(b, b')$ are such that 
\begin{eqnarray}
 \left[ \hat{A}, \hat{A'} \right]^{electrons} & = & 2i \hat{\sigma}_{\parallel} \sin(a'-a) \ , \nonumber \\
 \left[ \hat{A}, \hat{A'} \right]^{photons} & = & 2i \hat{\sigma}_{\parallel} \sin 2(a'-a) \ , \nonumber 
\end{eqnarray}
where $\sigma_{\parallel}$ is the Pauli matrix parallel to the direction of motion and the nature of the projection operator determines the argument of the sine.
When $\Delta a^{photon}=\pm 45^o$ \footnote{If Bob and Alice measure angles with the same convention, Bob must set his to $-22.0^o, -67.5^o$ if Alice uses $0^o,45^o$.}, %end footnote
 this local Bell operator's norm is consistent with the experimental data:  $ \left|S_{Bell} \right| \le 2\sqrt{2}$.

If we make the identification that the horizontally polarized state $|H\rangle$ corresponds to $|1,0\rangle$ with  $\sigma_z = +1$, and the vertically polarized state  $|V\rangle$ corresponds to $|0,1\rangle$ with $\sigma_z =-1$, then the operator $\sigma_{\parallel}$ corresponds to $\sigma_y$, assuming $\hat{A}$ and $\hat{B}$ measure mixtures of $\hat{\sigma}_z$ and $\hat{\sigma}_x$.

We can now ask what the matrix element evaluates to for unentangled and entangled photons.
  If Ou and Mandel's experiment~\cite{OU88} were performed without the beam splitter, or Weihs, {\it et al.}'s experiment~\cite{WZ98} sampled paired photons from anywhere on the two cones {\em except}  where they intersected, we would expect to find 
\begin{eqnarray}
\langle H_1V_2 \left| \hat{S}_{Bell}^2 \right| H_1V_2\rangle  % &=&
% 4 - (2i)^2 \sin(90^o)\sin(-90^o)  \langle H_1 \left| \sigma_{y}^{(1)} \right| H_1\rangle \langle V_2 \left| \sigma_{y}^{(2)} \right| V_2\rangle   , \nonumber \\
%  &=& 4 - 4\cdot 0\cdot 0 \ , \nonumber \\
  &=& 4 \ , \nonumber
 \end{eqnarray}
  since $\langle \sigma_z | \hat{\sigma}_y | \sigma_z\rangle = 0$.  This is  consistent with observation~\footnote{G. Weihs, {\it priv. comm.}, June 2004}.
  
   If we take the entangled  singlet wavefunction, $|\psi_e\rangle = \frac{1}{\sqrt{2}}| (H_1V_2 - V_1 H_2)\rangle$ then our result is
 \begin{eqnarray}
\langle \psi_e \left| \hat{S}_{Bell}^2 \right| \psi_e\rangle   %&=&
% 4 -4 \cdot \frac{1}{2}
% \langle H_1V_2 - V_1 H_2\left| \sigma_y^{(1)} \sigma_y^{(2)} \right|H_1V_2 - V_1 H_2\rangle   , \nonumber \\
%  &=& 4 - 4  \cdot \frac{1}{2} (- 2 \cdot i\cdot -i) , \nonumber \\
  & = & 8 \ ,\nonumber
 \end{eqnarray}  
  due to non-zero cross-terms ({\it e.g.} $\langle H\left| \hat{\sigma}_y \right| V\rangle = i$).  This is also consistent with observation~\cite{OU88, WZ98}.

%If we take unentangled but generic elliptically polarized photons, $| C(\phi)\rangle = \cos\phi|H\rangle + i \sin\phi |V\rangle$,  we still have cross terms and we still have a large bound on the Bell operator
%\begin{eqnarray} 
%\langle C^1 C^2_{\perp} \left| \hat{S}_{B}^2 \right|  C^1 C^2_{\perp}\rangle &=&
% 4 -  4  \langle C^1 \left| \hat{ \sigma}_y^{1} \right|C^1 \rangle 
%    \langle C^2_{\perp} \left|  \hat{ \sigma}_y^{2} \right|C^2_{\perp} \rangle \nonumber \\
%    &=& 4 + 4 \sin 2\Delta a \, \sin 2\Delta b \, \sin^2 \phi \ , \nonumber \\
%    &=&  8   \ , \nonumber
%\end{eqnarray}
%for circular polarization ($\phi=\pi/4$), a somewhat surprising result. Unfortunately, $\langle \hat{A}\hat{B} \rangle = \langle \hat{A} \rangle \langle \hat{B} \rangle$,  and $\langle C|  \hat{ A} |C \rangle = \cos 2\phi \cdot \cos 2\theta_a$, which implies $\langle \hat{S} \rangle = \cos^2 2\phi \sqrt{2} = 0$ for the circularly polarized case with the usual choice of detector angles.

Let us now consider the case of Ou and Mandel~\cite{OU88}, who used a second quantized (QED) wavefunction  to describe their data (where $|1_{dp}\rangle \equiv |N^{photon}_{detector, polarization}\rangle$),
\begin{eqnarray}
|\psi\rangle &=& (T_x T_y)^{1/2} |1_{1x},1_{2y}\rangle
                    +(R_x R_y)^{1/2} |1_{1y},1_{2x}\rangle \nonumber \\
                & &    -i(R_y T_x)^{1/2} |1_{1x},1_{1y}\rangle
                    +i(R_x T_y)^{1/2} |1_{2x},1_{2y}\rangle  \ , \nonumber
\end{eqnarray}
and creation-annihilation operators to build a joint intensity operator.  Then, even though the wavefunction can be rewritten in a factorized manner,
\begin{eqnarray}
|\psi\rangle    &\equiv&
                 ( \sqrt{T_x} | 1_{1x}\rangle + i\sqrt{R_x}|1_{2x}\rangle ) \nonumber \\
            & &         ( \sqrt{T_y} | 1_{2y}\rangle - i\sqrt{R_y}|1_{1y}\rangle ) 
               =  |\psi_x\rangle |\psi_y\rangle  \ . \label{eq:factord}         
\end{eqnarray}
the coincidence operator $E_1^{(-)}E_2^{(-)}E_1^{(+)}E_2^{(+)}$, where
\begin{eqnarray}
E_1^{(+)}&=&\cos \theta_a \sqrt{T_x} \hat{a}_x^{\dag} -i\sin\theta_a \sqrt{R_y}\hat{a}_y^{\dag} \ , \nonumber \\
E_2^{(+)}&=&i \cos \theta_b \sqrt{R_x} \hat{a}_x^{\dag} + \sin\theta_b \sqrt{T_y}\hat{a}_y^{\dag} \ , \nonumber
\end{eqnarray}
 will ensure that both $|\langle \hat{S}\rangle| \sim 2\sqrt{2}$ and $\langle \hat{S}^2\rangle \sim 8$.  (The components of the wavefunction that would send two particles to Alice or Bob do not contribute to coincidences.)  Since this wavefunction is of a product form, a de Broglie-Bohm~\cite{holland}  model for each photon would result in local forces and trajectories.

\section{Conclusions}

The implicit assumption of   simultaneous measurability or temporal order-independence of   measurements  at different orientations means that Bell's claim of  a  {\em locality}  bound is also a time independence or Heisenberg {\em irrelevance}  constraint .   {\em Classical} local hidden variable theories are precluded by experiment, but  {\em non-classical} ({\it i.e.} non-commutative), dynamic local hidden variable theories are not subject to Bell's original limit, but to the weaker Cirel'son~\cite{C80} limit of $2\sqrt{2}$; nonlocality presumably only shows up with the violation of that weaker limit.  The additional terms of    equations~\ref{eq:gallupslimit} or~\ref{eq:bellsquare} only contribute if non-classical effects occur locally at both locations; they do not require a distant particle to instantaneously affect a nearby particle's behavior in any way. 

 To the extent that the Copenhagen interpretation ($\lambda_i \equiv \psi(t_i)$ in the derivations) and the de Broglie-Bohm interpretation ($\lambda_i = \psi(t_i)$+hidden variables) both give the same answers in all experimental situations, our result shows that neither interpretation {\em needs} to be non-local in order to explain the data.

%\footnote{
% If the EPR argument were to be rephrased, Alice measures a component of spin with $a=0^o$ and Bob measures a component with $b=-90^o$; let us assume that $A(a)\equiv \sigma_{1z}  = +1$ and $B(b)\equiv \sigma_{2x}=+1$.  From the nature of the singlet state, EPR would {\em infer} a $``\sigma_{1x}"=-1$ and $``\sigma_{2z}"=-1$ and conclude that if a measurement value can be inferred (``with unit probability''), then there must be an element of reality that corresponds to it: Alice's  $\lambda_{1x}$ and $\lambda_{1z}$ must {\em both} exist.  Such inferences do not require one to perform the counterfactual experiments $A(a'=90^o)$ or $B(b'=0^o)$ at the same time and place as the actual $A(a)$ and $B(b)$,  nor is there any presumption that because $\lambda_z(t_0)$ resulted in $B(b,t_0)=+1$ that $\lambda_x(t_>)$ would still result in $B(b',t_>)=-1$ when measured later.  
 
% $\lambda_x$ and $\lambda_z$ both lie at the end of (different) long chains of  inference (and they are still ``hidden'') while Bell's derivation requires $B(b)$ and $B(b')$ be treated on the same footing, not one measured, one inferred; by introducing time into our derivation, we have put the measurements on that same (factual) level.
 %}

  For a single particle, the de Broglie-Bohm interpretation is a non-classical, dynamic  hidden variable theory.  It shows how  variables  evolve  in a manner that makes it clear why they can't be measured simultaneously or give the same result if measured in a different order.  If applied to a factorizable wavefunction of multiple particles, it is also a {\em local} HVT.

For the case of entangled particles that violate Bell's inequality,   the entanglement is a result of  the ``post-selection'' of only coincidences between Alice and Bob.   Ou and Mandel's  wavefunction will generate local dBB trajectories for each particle, showing that dBB is non-problematic even for multiple ``entangled'' particles.

  Similar arguments should apply to experiments like that of Weihs, {\it et al.}~\cite{WZ98}, where the entanglement is ``pre-selected'' before being fed into the optical fibers (throwing away the $\sim 99 \%$ of the photons in the non-intersecting parts of the down-conversion cones).  
  
  Using a ``post-selection'' wavefunction, as is usual in first quantization analyses, dBB will generate non-local forces on each trajectory, but each trajectory will contribute to a coincidence event; using a factorized ``pre-selection'' wavefunction appropriate to second quantization will generate {\em local} forces along each trajectory, but only some trajectories will satisfy the coincidence conditions.  A model {\em may} be nonlocal, but it doesn't {\em have} to be nonlocal to explain the data.  Computational efficiency hardly seems like an adequate reason to give up locality.

If Bell's  inequality depends on  $\lnot HUP$ and $LOC$, then violation of his inequality requires us to accept $HUP$ or $\lnot LOC$.  Nonlocality is not a choice, however, since our new derivation, which assumes $HUP$ and $LOC$, generates a weaker inequality that is respected by all experiments.

% It seems that the experiments  to date have  only shown that both Alice's and Bob's measurements are constrained by that {\em local} quantum effect, Heisenberg's Uncertainty Principle.  

%\newpage

\bibliography{bell3}% Produces the bibliography via BibTeX.

\end{document}